\documentclass[a4paper]{article}
\usepackage{amsfonts}
\usepackage{amsbsy}
\usepackage{amsmath}

\newcommand{\beq}{\begin{equation}}
\newcommand{\eeq}{\end{equation}}

\newcommand{\p}{\mathbb P}

\begin{document}

\title{\bf{Multiple-event probability in general-relativistic quantum mechanics: a discrete model}}

\author{\large Mauricio Mondragon, Alejandro Perez, Carlo Rovelli
 \\[1mm]
\em \small{
Centre de Physique Th\'eorique de Luminy%
\footnote{Unit\'e mixte de recherche (UMR 6207) du CNRS et des
Universit\'ees de Provence (Aix-Marseille I), de la Mediterran\'ee
(Aix-Marseille II) et du Sud (Toulon-Var); laboratoire affili\'e \`a
la FRUMAM (FR 2291).} , Universit\'e de la M\'editerran\'ee, F-13288
Marseille, EU}}

\date{\small\today}
\maketitle

\begin{abstract}

\noindent We introduce a simple quantum mechanical model in which
time and space are discrete and periodic.
These features avoid the complications related to
continuous-spectrum operators and infinite-norm states.
The model provides a tool for discussing
the probabilistic interpretation of generally-covariant quantum systems,
without  the confusion generated by
spurious infinities.
We use the model to illustrate
the formalism of general-relativistic quantum mechanics, and to test the
definition of multiple-event probability introduced in a companion paper.
We consider a version of the model with unitary time-evolution and a
version without unitary time-evolution.
\\
\\
PACS number(s): 04.20.Cv, 03.65.Ta, 04.60.Ds, 04.60.Pp
\end{abstract}

\section{Introduction}

Fundamental physical systems may be governed by a Wheeler-DeWitt-like
equation, rather than a Schr\"odinger equation.  For such systems, the
conventional probabilistic interpretation of quantum mechanics cannot be
directly applied, and the formulation of a consistent probabilistic interpretation
is a controversial  problem on which there is no
consensus yet \cite{Kiefer,Isham,Kuchar,Hartle:1992as,covariant,
book,Dittrich:2004cb,pw,Dolby:2004ak,olson}.
The complexity of this \emph{conceptual} problem is increased by specific \emph{technical}
difficulties related to the presence of continuous-spectrum operators,
infinite-norm states, gauge-groups having infinite volume, and
the like. These technical difficulties can often be
resolved by appropriate techniques, such as generalized states,
Gelfand triples, group integration, and similar.  But the technical difficulties
are often muddled with the core conceptual problem of making sense of
general-relativistic quantum mechanics, creating a confusing situation.
In this paper, we introduce a simple discrete
model where all interesting operators are
bounded and have discrete spectrum, so that many of these technical
difficulties do not appear. The model provides a tool for discussing
the probabilistic interpretation of generally-covariant quantum systems,
without getting entangled in the confusion generated by spurious
infinities and similar complications.

We use the model for illustrating the general-relativistic
formulation of quantum mechanics developed in \cite{covariant, book}
and in particular the prescription for computing  multiple-event
probability given in the companion paper  \cite{uno}. In \cite{uno},
we have argued that multi-event probability  can be always
re-conducted to single-event probability, provided that the dynamics
and the quantum nature of physical measuring apparatuses are taken
into account. Here we use the discrete model for illustrating and
testing this general technique for computing multiple-event
probabilities.

We begin by introducing the model in the simpler context in which
there is unitary evolution in time (Sec.\;\ref{model}). We discuss
its general-relativistic formulation in Sec.\;\ref{gr}. The
probabilistic interpretation is discussed in in Sec.\;\ref{sect:
multuntary}. The interesting ``timeless'' case, where there is no
unitary time-evolution is discussed in Sec.\;\ref{timeless}.  The
modification to the formalism required in the continuum case will be
discussed elsewhere \cite{continuum}.

As pointed out in \cite{uno}, the formalism that we obtain shows an
intriguing convergence  with the Hartle-Halliwell's history
formulation of quantum mechanics \cite{Hartle:1992as}. This
formulation of quantum theory has been  largely motivated by the
need of understanding the probabilistic interpretation of quantum
mechanics in the ``timeless" context, and the problems has been
discussed in depth; see \cite{Halliwell} and references therein. The
conceptual difference between the two approaches has been discussed
in detail in \cite{uno}.

\section{The model}\label{model}

Consider a particle in a finite region of two-dimensional
space-time. For simplicity, assume that space and time are both
periodic and the region has the topology of a torus. We shall later
take the time period to infinity. Discretize this region by means of
a lattice formed by  $N=N_x\times N_t$ points. Denote the lattice
points as $s=(x,t)$, where $x=1,2,...,N_x$ and $t=1,2,...,N_t$.  For
convenience, we take $x$ (respectively \,$t$) periodic modulo $N_x$
(respectively \,$N_t$) that is, we identify $x+N_x$ with $x$, and
$t+N_t$ with $t$. We write the space-time quantum mechanical wave
function of the system in the form $\psi(x,t)$, where, we insist,
$x$ and $t$ are here discrete, integer, variables.

To begin with, assume the dynamics is given by  a unitary evolution
in $t$.  We drop this assumption below in Sec.\;\ref{timeless}. Thus
assume for the  moment that there is a unitary $N_x\times N_x$
matrix $\cal U$ and the wave functions $\psi(x,t)$ that are
consistent with the dynamics of the theory are those satisfying \beq
            \psi(x,t+1) = {\cal U}\ \psi(x,t) = \sum_{y=1}^{N_x}\ \   {\cal U}_x{}^{y}\
            \psi(y,t).
            \label{dyn1}
\eeq This is a discrete version of the Schr\"odinger equation: we
can intuitively view the relation with a continuous theory as given
by ${\cal U}\sim e^{-iH_0\tau}$, where $H_0$ is the conventional
non-relativistic hamiltonian and $\tau$ is the lattice spacing in
the time direction (we put $\hbar=1$ throughout this work). In other
words, $\cal U$ is unitary-step time-evolution operator, that is,
the the operator that advance the state by one lattice step in the
time direction. The full time-evolution operator is given by \beq
            {\cal U}(t)= {\cal U}^t.
\eeq
Its matrix elements
\beq
W(x',t'; x,t) =  {\cal
U}_{x'}{}^x(t'-t) \label{W}
\eeq
form the discrete analog of the
propagator of the Schr\"odinger equation.  We assume that  $ {\cal
U}^{N_t}=1\!\! 1$, in order to respect the periodicity, a condition
that disappears in the $N_t \to \infty$ limit in which we are
interested. In non-relativistic
quantum mechanics, ${\cal U}_{x'}{}^x(t'-t)$ is interpreted as the
probability amplitude of finding the system at the point $x'$
at time $t'$ if we have found the system at the point $x$
at time $t$.

One may also consider a variant of the model, in
which ${\cal U}_x{}^{y}$ is non-vanishing only if $|x-y|\le k$,
where, say, $k=1$. This is typically the case when ${\cal
U}_x{}^{y}$ is obtained from the discretization of a Schr\"odinger
equation, and derivative operators are discretized by
first-neighborhood differences. We call this version ``micro-local",
since it propagates at finite speed on the lattice. (Recall, however, that
the nonrelativistic Schr\"odinger equation propagates at infinite speed.)
This micro-locality is in general inconsistent with the unitarity of ${\cal U}$.

\section{General relativistic formalism}\label{gr}

We now illustrate the general-relativistic formulation of the model
introduced above.  The \emph{kinematical} state space $\cal K$ is
the linear space formed by all complex functions $\psi(x,t)$,
whether or not they satisfy eq.\;\eqref{dyn1}. More precisely, we
define $\cal K$ as the Hilbert space ${\cal K}=\mathbb C^{N_x\times
N_t}$ spanned by the orthonormal basis $|s\rangle=|x, t \rangle$. We
write $\psi(x,t)=\langle x, t|\psi\rangle =\psi(s) =\langle
s|\psi\rangle$. The  \emph{physical} state space $\cal H$ is the
linear subspace of $\cal K$ formed by the states that satisfy the
dynamics, namely by the functions $\psi(x,t)$ that satisfy
(\ref{dyn1}). Since these states are uniquely determined by their
value on any fixed time slice $t=constant$, we have that ${\cal
H}\sim \mathbb C^{N_x}$.  Notice that, because of the simplicity of
our model, $\cal H$ is a \emph{proper} subspace of $\cal K$; this is
not true in general (see chapter 5 of \cite{book}).

Let us give a more covariant looking form to (\ref{dyn1}).   For
this purpose, let us introduce an $N\times N$ matrix $\cal T$
(recall $N$ is the total number of lattice points) whose only
non-vanishing matrix elements are \beq
           {\cal T }_{x't'}{}^{xt} = \delta_{t'}^{t+1}\  {\cal U}_{x'}{}^{x}.
            \label{covdyn}
\eeq  Notice that $\cal T$ is unitary and \beq {\cal T}^{N_t}=1\!\!1
\label{topolino} \eeq (here $1\!\!1$ is the unity operator over
$\mathcal{K}$). Then we can write the dynamical law (\ref{dyn1}) in
the covariant form ${\cal T}\psi=\psi$, that is \beq
            \sum_{s=1}^N\ {\cal T }_{s'}{}^{s} \ \psi(s)=\psi(s').
            \label{dwdw}
\eeq  This is the discrete analog of the continuous Wheeler-DeWitt
equation (see \cite{uno}).
 In other words,
we can intuitively view the relation with the continuous theory as
given by ${\cal T}\sim e^{-iH\tau}\sim e^{-i(p_t+H_0)\tau}$, where
$H$ is the relativistic hamiltonian \cite{book}, or ``hamiltonian constraint"
and $\tau$ is a small parameter-time step.

The generalized projector $\p$ that defines the relativistic
formalism is defined by
\beq
               \p= \sum_n\   {\cal T}^n,
               \label{defdis}
\eeq
where, because of the periodicity ${\cal T}^{n+N_t}={\cal T}^{n}$,
in (\ref{defdis}) we sum only over $n=1,...,N_t$.   The definition  (\ref{defdis})
can be seen as the discrete version of the expression
\begin{equation}
\p=\int
d\tau\ e^{-i\tau(p_t+H_0)},
\end{equation}
which defines the generalized projector in the continuous case
(see eq.\;(5.58) of \cite{book} and eq.\;(5) of \cite{uno}. This expression
is also related to the ``group averaging" technique, see \cite{Marolf:2000iq}).

It is easy to check that $\p$ is $N_t$ times the projector operator
$\tilde \p: {\cal K} \to {\cal H}$. In fact, if $\psi$ is in $\cal
H$, namely satisfies (\ref{dwdw}), we have immediately
$\p\psi=N_t\psi$. On the other hand, since $\cal T$ is unitary, it
is diagonalizable and its eigenvalues have the form $e^{i\alpha}$.
The physical Hilbert space  $\cal H$ is given by the eigenspace
associate with the eigenvalue $\alpha=0$: Because of the periodicity
(\ref{topolino}), $N_t\alpha=2\pi I$ for some integer $I$.  Acting
on the corresponding eigenstate, $\Phi_\alpha$ \beq \p\Phi_\alpha=
\sum_n\ {\cal T}^n\Phi_\alpha=
 \sum_n\   e^{in\alpha}\Phi_\alpha=
 \sum_n\   e^{in 2\pi I/N_t}\Phi_\alpha=0, \ \forall \ \alpha\neq 0\hspace{3em} QED.
\label{prova} \eeq

It is important to notice that while the projector $\tilde \p$ is
ill-defined in the $N_t \to \infty$ limit, the generalized projector
$\p$ remains well defined.  It is for this reason that $\p$ remains
meaningful also when $\cal H$ fails to be a proper subspace of $\cal K$.
 A straightforward calculation shows also
that
\beq
            \langle x', t' |\p|x, t \rangle =  {(\mathcal{U}^{(t'-t)})_{x'}}^x=  {(\mathcal{U}(t'-t))_{x'}}^x
            = {\cal U}_{x'}{}^{x}(t'-t) ;
\label{amplitude}
\eeq
that is: the matrix elements of $\p$ are the
matrix elements of the evolution operator or, what is the same, the
propagator (cf. eq. \eqref{W}).

Notice that we can write
 \beq
 \label{histories}
            \langle x', t' |\p|x, t \rangle =   ({\cal U}^{(t'-t)})_{x'}{}^{x}
            =\sum_{x_1...x_{t'\!-t-1}}   {\cal U}_{x'}{}^{x_1} {\cal U}_{x_1}
            {}^{x_2}\cdots {\cal U}_{x_{t'\!-t-1}}{}^{x}
            \equiv \sum_{\gamma_{(x,t){\Rightarrow} (x',t')}}
            U[\gamma]\, ,
 \eeq
where the sum is over all the ``histories" (sequences of lattice points)
$\gamma=\{(x_1,t),(x_2,t+1),....,(x',t')\}$ that start at the point
$(x,t)$, increase monotonically in time by one time unit at every
step, and end at the point $(x',t')$. This expression illustrates
the connection with the sum-over-history formalism
(see \cite{Hartle:1992as} and references therein). It is important to
observe that we can equivalently write
\beq
            \langle s' |\p|s \rangle
            =\sum_{s,s_1,s_2,...,s_n}   {\cal T}_s{}^{s_1} {\cal T}_{s_1}{}^{s_2}\cdots {\cal T}_{s_{n}}{}^{s'}
            \equiv \sum_{\gamma_{s {\Rightarrow} s'}}   U[\gamma]\,
            ,
\label{histories2}
\eeq
where now the sum can be extended over
\emph{all} possible histories. In fact, because of the form
(\ref{covdyn}) of the matrix $\cal T$, the only histories that have
nonvanishing amplitude are the ones that  increase monotonically in
time by one time unit at every step. Using this language, the fact
that the histories summed over increase monotonically in $t$ is not
an intrinsic property of the formalism, but only a property of the
matrix ${\cal T}_{s'}{}^{s}$. This is the property that we will drop
in Sec. \ref{timeless}.

This history language becomes particularly
 transparent in the micro-local case, in which
the transfer matrix $\cal U$ is directly given by a (non-unitary)
discretization of the Schr\"odinger equation. Say, for instance,
that $H_0=\frac{1}{2m}p^2+V(x)$, then, approximating the derivative
with finite differences, we write the Schr\"odinger equation
(\ref{dyn1}) in the form \beq \psi(x,t+1) = \alpha\, \psi(x+1,t) +
\alpha\, \psi(x-1,t) +  \beta(x)\, \psi(x,t)
            \label{dyn}
\eeq (if $a$ and $\tau$ are the spatial and temporal the
lattice-spaces, $\alpha=\frac{i\hbar\tau}{2ma^2}$ and
$\beta(x)=1-\frac{
i\tau}{\hbar}\big(\frac{\hbar^2m}{a^2}+V(x)\big)$).
Then the only non-vanishing matrix elements of the $\cal T$ matrix
are \beq {\cal T}_{x't'}{}^{xt} =\delta_{t'}^{t+1}\ \left(\alpha\,
(\delta_{x'}^{x+1} +\delta_{x'}^{x-1})+\beta(x) \,
\delta_{x'}^x\right). \eeq That is, $\cal T$ has non--vanishing
elements only for histories that are ``continuous" on the lattice,
namely formed by sequences of adjacent (or diagonal) lattice sites.

Let us now come to an important point.  For the moment, we have
introduced $\cal H$ as a linear space, without saying anything
about its Hilbert-space structure, namely about its scalar product.
As a proper subspace of $\cal K$, the space $\cal H$ inherits a scalar product
from $\cal K$.  However, it is far more convenient to define a
\emph{different} scalar product in $\cal H$, defined as the
scalar product in $\cal K$,  \emph{divided by $N_t$}. That is, if
$\psi,\psi'\in {\cal H}$, we define
\beq
\langle \psi' |\psi \rangle_{\cal H} \equiv \frac{1}{N_t}\ \langle
\psi' |\psi  \rangle_{\cal K}.
\label{sp}
\eeq
This definition follows easily from the general definition of the scalar product
of a physical space (eq. (5.64) and following lines in \cite{book},
and eq. (4) in \cite{uno})
\beq
\langle s' |s \rangle_{\cal H} \equiv \langle \p
s' |\p s \rangle_{\cal H}\equiv  \langle s' |\p | s \rangle_{\cal
K}.
\eeq
Thus, we promote $\cal H$ to an Hilbert space by defining its
scalar product by (\ref{sp}).  The advantage of this, of course,
is that in the limit $N_t\to\infty$ the scalar product $\langle
\ | \ \rangle_{\cal H}$ remains well defined. In other words,
this Hilbert structure is well defined even when $\cal H$
is not a proper subspace of $\cal K$.
Notice that it follows from this definition that $|x, t \rangle$ is a
normalized vector in $\cal K$ and also $\p|x, t \rangle$ is a
normalized vector in $\cal H$.

From now on, we take $N_t \to
\infty$. We interpret this limit as describing the infinite time
evolution, not the continuous limit of the lattice spacing.

\section{Probabilistic interpretation}\label{sect: multuntary}

According to the basic postulate of relativistic quantum mechanics
\cite{book}, states in $\cal K$ represent possible observations. In
particular, the state $|x,t\rangle\in {\cal K}$ represents the event
$(x,t)$ in spacetime: say ``the particle is detected by an
apparatus" at the spacetime point $(x,t)$. Then (see eqs.
(5.59)-(5.60) in \cite{book}, or eq. (7) in \cite{uno}) the
probability amplitude for observing the event $(x', t')$ if the
event $(x, t)$ was detected, is given by \beq
 \mathcal{A}_{(x,t){\Rightarrow} (x',t')}= \langle x', t' |\p|x, t \rangle.
\eeq By equations (\ref{W}) and (\ref{amplitude}), this is the
conventional amplitude of standard quantum mechanics, that is, the
propagator. The interest of this formulation, on the other hand, is
that it remains meaningful and well defined when the dynamics is not
unitary in time.

The above definition of probability amplitude is sufficient to
define the probability of single events, described by individual
states in $\cal K$. The problem left open in \cite{book}, and
addressed in \cite{uno}, was to define the probability for
\emph{multiple} events, or for a subspace of $\cal K$ with dimension
larger than 1.   This cannot be avoided, in particular, when dealing
with operators with continuous spectrum, because physical
apparatuses have always finite resolution and therefore in general
cannot detect individual events. Let us therefore move to
multiple-event probabilities.

A case that does not raise difficulties is to define the probability that the
particle be detected at $(x,t)$ OR $(y,t)$, where $x\ne y$, but,
notice, the $t$ of the two events is the same.  In this case, the
two events are orthogonal in $\cal H$, because
\beq
            \langle y, t |\p|x, t \rangle =   ({\cal U}(0))_y{}^{x}=\delta_y{}^{x}.
\eeq We are therefore in the simple case (i) of Sec.\,III of
\cite{uno}, where, following  \cite{uno}, we postulate that
probabilities add.  This is of course nothing but the standard
assumptions in non-relativistic  quantum mechanics demanding that
the probability of being at any one of two different points {\em at
the same time} is the sum of the two probabilities of being at the
individual points.   Let us move on to the more interesting case.

What is the probability that the particle be detected at $(y,t_1)$
OR $(y,t_2)$, where, say, $t_1<t_2$?  For simplicity, let us focus,
instead, on the probability for the particle to be detected in
$(y,t_1)$ AND $(y,t_2)$; the two are easily related (see eq. (14) in
\cite{uno}). Now we have \beq
            \langle y, t_2|\p|y, t_1 \rangle =  { \cal U}_y{}^{y}(t_2-t_1)
\eeq which in general does not vanish. This is  case
(ii) of Sec.\;III of  \cite{uno}.   According to \cite{uno}, this
probability can be given a physical meaning \emph{only} by
explicitly coupling the
system to an apparatus that interacts with the system at time $t_1$
at the point $y$, namely at the space-time point $s_1=(y,t_1)$.  Let
us therefore do so.

We model the apparatus in the simplest possible form.
Introduce a two-state system $|A\rangle$ with $A=Yes, No$,
and couple it to our system.  The kinematical state of the combined
system is  ${\cal K}_{ext}=\mathbb C^{N_x\times N_t}\otimes \mathbb
C^2$, spanned by the orthonormal basis $|\sigma\rangle=|x,t,A
\rangle=|s,A \rangle$. We define the dynamics by replacing the
operator $\cal T$ with the operator ${\cal T}_{s_1}={\cal T}_{x_1t_1}$
defined by
\beq
({\cal T}_{s_1})_{\sigma'}{}^{\sigma}={\cal
T}_{s'A'}{}^{sA}= {\cal T}_{s'}{}^{s}\ {\delta}_{A'}^A +
\delta_{s_1}^s {\cal T}_{s'}{}^{s} ({\cal I}_{A'}{}^A
-{\delta}_{A'}^A ). \label{coupling}
 \eeq
The first term is the free evolution of the system and the free
(trivial) evolution of the apparatus.  The second term is the
interaction.  It is different from zero only at the space-time point
$s_1=(x_1,t_1)$, that is, the apparatus flips, if and only if the
particle pass through $x=x_1$ at time $t=t_1$. Here, the evolution
of the apparatus is given, instead than the trivial evolution
operator ${\delta}_{A'}^A$, by the matrix \beq
{\cal I}_{A'}{}^A \equiv \left(\begin{array}{cc} 0& 1\\
1 & 0\end{array} \right),
\label{defA}
\eeq
which flips the state of the apparatus.  We call
$\p_{s_1}=\p_{x_1,t_1}$ the generalized propagator defined by
this coupling, namely by inserting the operator ${\cal T}_{x_1t_1}$
defined in (\ref{coupling}) into (\ref{histories2}). That is
\beq
 \langle s' |\p_{s_1}|s \rangle
=\sum_{\tilde s_1,\tilde s_2,...,\tilde s_n}   ({\cal
T}_{s_1})_s{}^{\tilde s_1} ({\cal T}_{s_1})_{\tilde s_1}{}^{\tilde
s_2}\cdots ({\cal T}_{s_1})_{\tilde s_{n}}{}^{s'} \equiv
\sum_{\gamma_{s {\Rightarrow} s'}} U_{s_1}[\gamma]\,
.\label{histories3} \eeq  It is easy to see that this implies
\beq\label{eccolo0} \langle x', t', A' |\p_{x_1,t_1}| x, t, A
\rangle = {\cal U}_{x'}{}^{x}(t'-t)\ {\delta}_{A'A} \eeq if $t$ and
$t'$ are both smaller or larger than $t_1$ (the interaction time),
or if $t'<t_1<t$; and \beq
 \langle   x', t', A'  |\p_{x_1,t_1}|  x, t, A \rangle =
\sum_{x''} {\cal U}_{x'}{}^{x''}(t'-t_1) \ \Big( {\delta}_{A'A} +
\delta_{x''}^{x_1} ({\cal I}_{A'A}- {\delta}_{A'A}) \Big) {\cal
U}_{x''}{}^{x}(t_1-t) \eeq if $t<t_1<t'$. Notice that this
interaction is designed for the apparatus to flip state if and only
if the particle is at $x_1$ at time $t_1$.  Using (\ref{histories}),
this can be rewritten in a more covariant looking form \beq
            \langle x', t', A' |\p_{x_1,t_1}|x, t, A \rangle
            = \sum_{\gamma_{(x, t){\Rightarrow} (x',t')}}   U[\gamma]\
            \left({\delta}_{AA'} + \ \delta^\gamma_{(x_1,t_1) } \ ({\cal I}_{AA'} -{\delta}_{AA'} )
          \right)
\label{eccolo} \eeq where $\delta^\gamma_{(x_1,t_1) }=1$ if the
history goes through $x_1,t_1$ and $\delta^\gamma_{(x_1,t_1) }=0$
otherwise.

The multiple probability amplitude is then given by (according to
Sec. IV.C of \cite{uno}) \beq {\cal
A}_{(x,t){\Rightarrow}((x_1,t_1),(x',t'))} = \langle  x', t', Yes
|\p_{x_1,t_1}| x, t,  No \rangle. \label{multipleprob} \eeq A
straightforward computation gives the probability (eq. (10) of
\cite{uno})\beq \mathcal{P}_{(x,t){\Rightarrow}((x_1,t_1),(x',t'))}
= |{\cal A}_{(x,t){\Rightarrow}((x_1,t_1),(x',t'))} |^2 = |{\cal
U}_{x'}{}^{x_1}(t'-t_1)|^2 \ \ |{\cal U}_{x_1}{}^{x}(t_1-t)|^2 \, .
\eeq The reader can check with a tedious but straightforward
calculation,
using standard quantum mechanics and the wave function collapse
algorithm,
 that this is the correct probability for this sequence of
measurements, namely
 for detecting the particle at  $x'$ at time
$t'$ \emph{and}  at $x_1$ at time $t_1$ if the particle was
initially at $x$ at time $t$ (assuming $t<t_1<t'$).

The coupling to an apparatus can be easily generalized to an
arbitrary number of apparatuses.   For instance, say we have
apparatuses at the space-time points $s_1,...,s_K$ coupled as
follows \beq {\cal T}_{s'A_1'...A_K'}{}^{sA_1...A_K}= {\cal
T}_{s'}{}^{s}\ {\delta}_{A_1'}^{A_1}\cdots \ {\delta}_{A_K'}^{A_K} +
\sum_{k=1,K} \delta_{s_k}^s {\cal T}_{s'}{}^{s} \
{\delta}_{A_1'}^{A_1}\cdots {\delta}_{A'_{k-1}}^{A_{k-1}}({\cal
I}_{A_k'}{}^{A_k} -{\delta}_{A_k'}^{A_k} )
{\delta}_{A'_{k+1}}^{A_{k+1}} \cdots \ {\delta}_{A_K'}^{A_K}.
\label{couplingmulti} \eeq Then the probability to observe the
system at the space-time point $(x',t')$ in conjunction with all
apparatuses having detected the particle at all the events $s_1,s_2,
\ldots, s_K$, if the system was observed at the space-time point
$(x,t)$ in conjunction with all apparatuses being in the $No$ state
is
\beq
\mathcal{P}_{(x,t){\Rightarrow}((x_1,t_1)...,(x_K,t_K),(x',t'))} = |\langle
x',  t', \underbrace{Yes,..., Yes}_{\mbox{$K$}} |\p_{s_1...s_K}| x,
t, \underbrace{No, ..., No}_{\mbox{K}} \rangle|^2
\label{multipleprob2}
\eeq
The possibility of taking the
continuum limit of this expression will be discussed in \cite{continuum}.
Again, the reader can check  that this is the correct probability for a
sequence of measurements, where the wave function is repeatedly collapsed at each
measurement.

What we have done so far, indeed, is nothing more than a reformulation
of the standard probabilistic interpretation of quantum theory, including
the probabilities computed by using the collapse postulate.  What is
interesting is that \emph{this} formulation can be readily generalized to the
case where there is no unitary evolution, and therefore no evident
time-ordering in the system.   In fact, the reader should notice that
the expression (\ref{multipleprob2})   does not depend
\emph{explicitly} on the time-ordering of the measurements.
Time-ordering is implicitly implemented by the \emph{dynamics} of
the coupled system, which is coded into the form of $\cal T$. In other words, if
$s_2$ is earlier in time than $s_1$, then the probability
(\ref{multipleprob2}) becomes automatically the one given by first
projecting on $s_2$ and then on $s_1$. This is particularly
transparent in the history language: the histories summed over are
only the ones correctly ordered in time {\em because the dynamical
operator $\cal T$ selects only these histories.}  In this sense, in
this formalism the time ordering of the measurements is absorbed
into the definition of the dynamics. It is this fact that allows us
to generalize the formalism to a genuinely ``timeless" situation,
namely a situation in which the dynamics is not given by unitary
evolution in $t$. This is what we do in the following section.

\section{Extension to the timeless case}\label{timeless}

We now consider the extension of the discrete model to the case
where there is {\em no} unitary evolution in time.   We take the
same kinematics as the discrete dynamics above, but  a different
dynamics. That is, we take again a lattice with lattice points
$s=(x,t)$, but now there will not be a unitary evolution in $t$. The
kinematical Hilbert space is always spanned by the orthonormal basis
$|s\rangle=|x,t\rangle$, and we still define the quantum dynamics by
the discrete Wheeler-DeWitt equation (\ref{dwdw}) and by the
generalized projector (\ref{defdis}).

The novelty is that now we \emph{do not} assume $\cal T$ to have
the form (\ref{covdyn}).  This form, in fact, is
the discrete analog of $H=p_t+H_0$, which defines
the non-relativistic special case of generalized
quantum mechanics.

A particularly interesting case is given when $\cal T$ is still
micro-local, that is, has non-vanishing matrix elements ${{\cal
T}_{s'}}^{s}$ only for lattice points $s=(x,t)$ and $s'=(x',t)$ that
differ at most by $k$ lattice steps, namely for $|x'-x|<k$ and
$|t'-t|<k$, for instance with $k=1$. This property can follow from
instance from a discretization of a continuous relativistic
hamiltonian $H$ that has finite degree in the derivatives.

We assume for simplicity that the system is periodic, and there is
an integer $Z$ such that ${\cal T}^Z=1\!\!1$. This does not need to
be a periodicity in one of the coordinates of the extended
configuration space. The example that follows, for instance, has a
periodic dynamics. Then the generalized projector is defined by \beq
  \p = \sum_{n=1,Z}\ {\cal T}^n.
\eeq
(Otherwise, we have to take the limit $
  \p = \lim_{Z{\Rightarrow}\infty} \frac{1}{Z} \sum_{n=1,Z}\ {\cal T}^n.$)
All expressions of the previous section continue to hold. In
particular, we can still use the coupling (\ref{coupling}) to the
apparatus system, and define probabilities amplitudes by
(\ref{multipleprob}). This shows that multiple probabilities can be
defined in general also in a system without unitary time evolution.

The physical interpretation of these probabilities is discussed in detail
in the companion paper \cite{uno}, and will not be repeated here. We
only recall here the important fact that in the general case, the probability
of detecting the event $s$ is \emph{not} interpreted as the number  of times
the system is detected in  $s$, {\em instead of being detected in other
positions at equal time}.  Rather, it is interpreted as the number  of times
the system is detected in  $s$, {\em instead of not being detected}.  For instance, the number of time a detector at $s$ will click,
divided the total number of runs of the experiment.   The two definitions
are equivalent in the non-relativistic case, because in this case if
a particle is not at $x$, then the particle must be in another one (and only
one other) position of the equal time surface; in the timeless case
the first definition becomes meaningless, but the second remain
meaningful.

As a concrete example, consider a discretized version of the finite
dimensional model studied in \cite{daniele}.  In the continuum, this
model is defined by a two-dimensional extended configuration space,
with coordinates $a$ and $b$, and by the relativistic hamiltonian
\beq
    H=\frac{1}{2}\left(p_a^2+p_b^2+a^2+b^2\right)-E\, ,
\eeq
where $E$ is an integer constant. We refer to
\cite{daniele} for all details and motivations. The motions are
closed lines in the extended configurations space (in fact,
ellipses), and therefore this model does not admit a conventional
hamiltonian formulation (because the motion of a hamiltonian system
are always monotonic in $t$ and therefore cannot be closed lines).
The dynamics of the model is determined by the dynamics
of two harmonic oscillators (with $m=\omega=1$) with fixed total
energy $E_a+E_b=E$. (The difference is that here the time variable
is assumed to be unobservable.) In fact, the two commuting operators $
H_{a}=\frac{1}{2}\left(p_a^2+a^2\right)$ and
$H_{b}=\frac{1}{2}\left(p_b^2+b^2\right)$ are diagonalized by the
states \beq \Psi_{n_an_b}(a,b)=\langle a, b|n_a,n_b \rangle =
H_{n_a}(a)H_{n_b}(b) \eeq where $n_a,n_b=0,1,...,\infty$ and
$H_n(x)$ are the harmonic oscillator eigenstates. Of course: \beq
 |a,b \rangle = \sum_{n_a=0}^\infty \sum_{n_b=0}^\infty\
  \overline{H_{n_a}(a)} \ \overline{H_{n_b}(b)}\ |n_a,n_b \rangle.
  \label{infinito}
\eeq
The relativistic hamiltonian gives
\beq
H |n_a,n_b \rangle =  (E_a+E_b-E) |n_a,n_b \rangle =
 (n_a+n_b+1-E) |n_a,n_b \rangle.
\eeq
and defines the physical space via $H\psi=0$.
In order for  solutions to exist, $E$ must be integer and the physical
Hilbert space $\cal H$ is spanned by the $E$ states $|n,E-1-n \rangle, \ \ n=0
 ,..., E-1$. Here $\cal H$ is a proper subspace of $\cal K$. The generalized
 projector is precisely the projector
\beq
 \p = \frac{1}{2\pi}\int_0^{2\pi} d\phi\ \ e^{-iH\phi}=
 \sum_{n=0}^{E-1}\  |n,E-1-n \rangle\langle n,E-1-n|.
\eeq Its matrix elements are \beq \langle a',b'| \p|a,b\rangle =
\sum_{n=0}^{E-1} \ H_{n}(a')H_{E-1-n}(b') \ \overline{H_{n}(a)} \
\overline{H_{E-1-n}(b)}.
\eeq
For more details, see \cite{daniele}.

Let us discretize this model, in order to avoid probability
densities and distributions. This can be done as follows. Choose an
integer $N\gg E$. Cut-down the kinematical phase space to the space
spanned by the states where $|n_a,n_b \rangle, \ \
n_a,n_b=0,1,...,N-1$. We write $N=2M+1$ for later convenience.
Introduce the discretized ``position states" by analogy with
(\ref{infinito}), as
 \beq
 |a,b \rangle := \sum_{n_a=0}^{N-1}\sum_{n_b=0}^{N-1}  \
  \overline{H_{n_a}(a)}\ \overline{H_{n_b}(b)}\ |n_a,n_b \rangle
\eeq  where $a,b=-M\delta,-(M-1)\delta, ... , 0,
...,(M-1)\delta,M\delta$, and $\delta$ is a suitably chosen small
real number.\footnote{We want the lattice spacing to satisfy
$\delta\ll 1$ in order to be much smaller than the harmonic
oscillator vacuum state width $\sqrt{\hbar/m\omega}$, otherwise the
vacuum state lattice approximation would be very rough. In the
present case $\sqrt{\hbar/m\omega}=1$, since we have chosen units
where $\hbar=m=\omega=1$. We also want the lattice spacing to be
much smaller than the minimum allowed wavelength: the wavelength is
the length on which the state varies;  in order for the state to be
well approximated by a state on the lattice, the lattice
spacing must be much smaller. Since the maximum momentum allowed by
the constraint is $p_{max}\sim \sqrt{2E}$, this implies that
$\delta\ll\frac{2\pi \hbar}{p_{max}}$, that is, $\delta \ll
\frac{2\pi}{\sqrt{2E}}$. Finally, we choose $M$ such that the
(finite size) region covered by the lattice is much larger than the
maximum amplitude, that is, this region should be large enough to
accommodate any state. This implies $M \gg \sqrt{2E}/\delta$,
therefore $N\gg E$. Under these conditions, there are all reasons to
believe that the quantum dynamics of the discrete model approximates
well the continuum one.} Notice that $a$ and $b$ are here
\emph{discrete} variables.  The resulting discrete model can be taken as
a discrete approximation of the model studied in \cite{daniele}.

 The projector can
then be written in the discrete form
\beq
 \p =\sum_{n=0}^{E-1}\   |n,E-1-n \rangle\langle n,E-1-n|
 = \frac{1}{2N}\sum_{\tau=1}^{2N}  \ e^{-2\pi i H\tau/2N}.
\eeq  where $H |n_a,n_b \rangle=(n_a+n_b+1-E)\, |n_a,n_b \rangle$.
(Remember that $2N$ is the total number of kinematical states.) Equivalently
\beq
 \p =\frac{1}{2N}\sum_{\tau=1}^{2N}  \ {\cal T}^{\tau}
\eeq where \beq
 {\cal T} = e^{-2\pi iH/2N}
\eeq is the matrix giving the one-step parameter-time evolution in
the discretized model. Its matrix elements are
 \beq \label{Trel}
  {\cal T}_{a'b'}{}^{ab} =
  \langle a',b'|  {\cal T} |a,b \rangle
  =  \sum_{n_a=0}^{N-1}\sum_{n_b=0}^{N-1}
  H_{n_a}(a')\ H_{n_b}(b')\
e^{2\pi i(n_a+n_b+1-E)/2N}\ \overline{H_{n_a}(a)}\
\overline{H_{n_b}(b)}\, . \eeq Using this matrix, the matrix
elements of the projectors can be written as a sum over histories,
as in (\ref{histories2}), where, now, $s=(a,b)$ and we sum over all
histories long at most $2N$ steps.

We are now ready to couple the system to an apparatus and calculate
multiple-event probabilities. Let's say we have an apparatus that
detects the system at the point $(a_1,b_1)$. Then we modify $\cal T$
as follows
 \beq
 {\cal T}_{a'b'}{}^{ab}
\rightarrow ({\cal T}_{a_1b_1})_{a'b' A'}{}^{abA} = {\cal
T}_{a'b'}{}^{ab} \delta_{A'}^{A}+ \delta_{a^1b^1}^{ab}{\cal
T}_{a'b'}{}^{ab}\Big({\cal I}_{A'}^{A}-\delta_{A'}^{A}\big). \eeq
This gives the projector \beq \p_{a_1b_1} = \sum_{\tau=1}^{2N}\
({\cal T}_{a_1b_1})^{\tau} \eeq  and the probability amplitude of
finding the system at the point $(a',b')$ of the extended
configuration space, with the detector having detected the system in
the point $(a_1,b_1)$, if the system was detected at the point
$(a,b)$ with the detector in the $|No\rangle$ state, is \beq {\cal
A}_{(a,b)\Rightarrow((a_1,b_1),(a',b'))} = \langle a',b',Yes
|\p_{a_1,b_1}|a,b,No\rangle \eeq This is a well-defined
multiple-event probability in a system without unitary time
evolution. This shows that the prescription of \cite{uno} can be
implemented, at least in principle, also in a system without unitary
time evolution.

We close with a conjecture. It should be interesting to consider the
semiclassical limit of these probabilities. In \cite{daniele},
semiclassical coherent states corresponding to classical
trajectories (ellipses) were constructed.  Consider a semiclassical
kinematical wave packet peaked around the point $(a,b)$, and with
velocity $(p_a,p_b)$. This state is projected by $\p$ onto a a
semiclassical state peaked along the corresponding ellipses. In
\cite{daniele}, it was shown that the kinematical state space splits
into clockwise and anticlockwise moving states.  Let $\psi_{ab}^C$
and $\psi_{ab}^A$ be two such states, belonging to the clockwise and
anticlockwise subspace, respectively, both generating a
semiclassical motion along a given ellipses (but in opposite
directions). Let $(a,b)$,  $(a',b')$ and $(a_1,b_1)$ be three points
along these ellipses. Consider the two transition amplitudes \beq
\mathcal{A}_{\psi^C_{ab}{\Rightarrow}((a_1,b_1),(a',b'))} = \langle
a', b', Yes |\p_{a_1,b_1}| \psi^C_{ab}, No \rangle. \eeq and \beq
\mathcal{A}_{\psi^A_{ab}{\Rightarrow}((a_1,b_1),(a',b'))} = \langle
a', b', Yes |\p_{a_1,b_1}| \psi^A_{ab}, No \rangle. \eeq One may
expect that in the classical limit the first of these amplitudes is
suppressed
 if $(a,b){\Rightarrow}(a_1,b_1){\Rightarrow}$ $
(a',b')$ is ordered anti-clockwise, and the second is suppressed
 if $(a,b){\Rightarrow}(a_1,b_1){\Rightarrow}
(a',b')$ is ordered clockwise.

\section{Conclusion}

Using a simple discrete model, we have shown in detail how multi-event
probabilities can be computed in a ``timeless" quantum system,
governed by a Wheeler-DeWitt-like equation, instead than
a Schr\"odinger equation. We have computed these probabilities
using the strategy introduced in \cite{uno},
namely the observation that  multi-event probabilities can be
interpreted as probabilities for a single event, if the event includes the presence
of a ``record" of a feature of the system measured by an apparatus that has
interacted with the system itself.

In words, to say that\\
(i)  a particle has been detected at  $(x_1,t_1)$ \emph{and} at
$(x_2,t_2)$
can be interpreted as \emph{meaning} that we detect a state in which\\
(ii)  the particle is detected at  $(x_2,t_2)$ and there is a record that the particle has interacted
with another system at $(x_1,t_1)$.\\
This interpretation is conceptually satisfying, since how else could
we combine events happening at different times, if not by having
records? The key difference between (i) and (ii) is of course that
in conventional quantum mechanics the property of being at
$(x_1,t_1)$ and the property of being at  $(x_2,t_2)$ are
represented by quantum operators that do not commute.  While the
property of being detected at  $(x_2,t_2)$ and the existence of a
record that the particle has interacted with another system at
$(x_1,t_1)$ are expressed by operators that commute: hence there is
a single state where both properties are sharp.

Notice that nothing in the construction presented here assures us
that the probabilities computed are independent from the dynamics of
the apparatus and its coupling.  In fact, in general, the opposite
may very well be the case, especially in more complicated
situations, or in the continuous case. This is not a defect of the
formalism. To the contrary, it is an indication from the theory that
physically different apparatuses may give different results (see
\cite{Halliwell} and references therein). In other words, the
philosophy here is not that an abstract apparatus-independent
probability  has to be well defined. Rather, in any concrete
physical situation, where a given apparatus is present, the
formalism should be capable of providing the probabilities of the
observable outcomes.

\section*{Acknowledgments}

This work has been  partially completed thanks to the financial
support provided by the Programme Al{\ss}an, European Union
Programme of High Level (scholar-)Grants for Latin America, grant
No. E04D033873MX;  and by the Programme SFERE-CONACYT
(France-Mexico), grant No.\,166826.

\end{document}